\documentclass[twoside,11pt]{5ssmmp} %%% Please do not change this class !!! %%%
\usepackage{amsfonts}
\usepackage{amssymb}
\usepackage{amsmath}
\usepackage{fancyhdr}    % headings style
\usepackage{equations}
\usepackage{epic}
\usepackage{eepic}
\usepackage{epsf}
\usepackage{graphicx}
\usepackage{rotating}
\newcommand\rf[1]{(\ref{eq:#1})}
\newcommand\lab[1]{\label{eq:#1}}
\newcommand\nonu{\nonumber}
\newcommand\br{\begin{eqnarray}}
\newcommand\er{\end{eqnarray}}
\newcommand\be{\begin{equation}}
\newcommand\ee{\end{equation}}

\newcommand\lb{\lbrack}
\newcommand\rb{\rbrack}
\newcommand\llangle{\left\langle}
\newcommand\rrangle{\right\rangle}

\newcommand\llb{\left\lbrack}
\newcommand\rrb{\right\rbrack}

\renewcommand\({\left(}
\renewcommand\){\right)}
\newcommand\bv{\bigm\vert}               %%
\newcommand\bgv{\bigg\vert}              %%

\newcommand\bc{\begin{center}}
\newcommand\ec{\end{center}}

\newcommand\partder[2]{\frac{{\partial {#1}}}{{\partial {#2}}}}
\renewcommand\b{\beta}

\renewcommand\d{\delta}

\newcommand\eps{\epsilon}
\newcommand\vareps{\varepsilon}
\newcommand\g{\gamma}
\newcommand\G{\Gamma}

\newcommand\h{\frac{1}{2}}
\renewcommand\k{\kappa}
\renewcommand\l{\lambda}
\renewcommand\L{\Lambda}
\newcommand\m{\mu}
\newcommand\n{\nu}

\newcommand\vp{\varphi}

\newcommand\pa{\partial}

\newcommand\s{\sigma}
\renewcommand\S{\Sigma}
\renewcommand\t{\tau}
\renewcommand\th{\theta}

\newcommand\wti{\widetilde}
\newcommand\cA{{\mathcal A}}

\newcommand\cE{{\mathcal E}}
\newcommand\cF{{\mathcal F}}

\newcommand\cJ{{\mathcal J}}

\newcommand\cV{{\mathcal V}}

\newcommand{\ct}[1]{\cite{#1}}
\newcommand{\bib}[1]{\bibitem{#1}}

\newcommand\PRL[3]{\textsl{Phys. Rev. Lett.} \textbf{#1} (#2) #3}
\newcommand\NPB[3]{\textsl{Nucl. Phys.} \textbf{B#1} (#2) #3}

\newcommand\PRD[3]{\textsl{Phys. Rev.} \textbf{D#1} (#2) #3}

\newcommand\PLB[3]{\textsl{Phys. Lett.} \textbf{#1B} (#2) #3}
\newcommand\CQG[3]{\textsl{Class. Quantum Grav.} \textbf{#1} (#2) #3}

\newcommand\IJMPA[3]{\textsl{Int. J. Mod. Phys.} \textbf{A#1} (#2) #3}

\newcommand\Xdot{\stackrel{.}{X}}
\newcommand\xdot{\stackrel{.}{x}}

\newcommand\udot{\stackrel{.}{u}}

\newcommand\etadot{\stackrel{.}{\eta}}
\begin{document}
 \baselineskip=11pt

\title{Space-Time Compactification, Non-Singular Black Holes, Wormholes and 
Braneworlds via Lightlike Branes\hspace{.25mm}\thanks{\,Work supported in
part by Bulgarian National Science Foundation grant DO 02-257}}
\author{\bf{Eduardo Guendelman and Alexander Kaganovich}\hspace{.25mm}\thanks{\,e-mail address:
guendel@bgu.ac.il, alexk@bgu.ac.il}
\\ \normalsize{Physics Department, Ben Gurion University of the Negev} \\
\normalsize{Beer Sheva, Israel} \vspace{2mm} \\ 
\bf{Emil Nissimov and Svetlana Pacheva}\hspace{.25mm}\thanks{\,e-mail
address: nissimov@inrne.bas.bg, svetlana@inrne.bas.bg}
\\ \normalsize{Institute for Nuclear Research and Nuclear Energy}\\
\normalsize{Bulgarian Academy of Sciences, Sofia, Bulgaria} }
% \vspace{2mm} \\ \bf{Third author}\hspace{.25mm}\thanks{\,e-mail
% address: third@5ssmmp.mp}
% \\ \normalsize{Address of the third author} }

\date{}

\maketitle

\begin{abstract}
We describe a concise general scheme for constructing solutions of 
Einstein-Maxwell-Kalb-Ramond gravity-matter system in bulk space-time interacting 
self-consistently with one or more (widely separated) codimension-one
electrically charged {\em lightlike} branes. The lightlike brane dynamics is 
explicitly given by manifestly reparametrization invariant world-volume actions.
We present several explicit classes of solutions with different physical
interpretation as wormhole-like space-times with one, two or more
``throats'', singularity-free black holes, brane worlds and space-times
undergoing a sequence of spontaneous compactification-decompactification
transitions.
\end{abstract}

% {\Large{Text of your article}}
%%%%%%%%%%%%%%%%%%%%%%%%%%%%%%%%%%%%%%%%%%%%%%%%%%%%%%%%%%%%%%%%%%
%%%%%%%%%%%%%%%%%%%%%%%%%%%%%%%%%%%%%%%%%%%%%%%%%%%%%%%%%%%%%
\section{Introduction}
Lightlike branes (\textsl{LL-branes} for short) are singular null (lightlike) 
hypersurfaces in Riemannian space-time which provide dynamical description of 
various physically important  phenomena in cosmology and astrophysics such as:
(i) impulsive lightlike signals arising in cataclysmic astrophysical events 
(supernovae, neutron star collisions) \ct{barrabes-hogan}; 
(ii) dynamics of horizons in black hole physics -- the so called ``membrane paradigm''
\ct{membrane-paradigm};
(iii) the thin-wall approach to domain walls coupled to 
gravity \ct{Israel-66,Barrabes-Israel,Dray-Hooft}.

More recently, the relevance of \textsl{LL-branes} in the context of
non-perturbative string theory has also been recognized \ct{nonperturb-string}.

Starting with the pioneering papers \ct{Israel-66,Barrabes-Israel,Dray-Hooft}
the \textsl{LL-branes} have been exclusively treated in a ``phenomenological'' 
manner, \textsl{i.e.}, without specifying an underlying Lagrangian dynamics from which
they may originate. On the other hand, in the last few years we have proposed in a series of papers 
\ct{LL-brane-main,inflation-all,our-WH,rot-WH} a new class of concise
manifestly reparametrization invariant world-volume Lagrangian actions, 
% among them -- {\em Weyl-conformally invariant} ones, 
providing a derivation from first principles of the \textsl{LL-brane} dynamics.
The following characteristic features of the new \textsl{LL-branes} drastically
distinguish them from ordinary Nambu-Goto branes: 

(a) They describe intrinsically lightlike modes, whereas Nambu-Goto branes describe
massive ones.

(b) The tension of the \textsl{LL-brane} arises as an {\em additional
degree of freedom}, whereas Nambu-Goto brane tension is a given
{\em ad hoc} constant. %%%%%%%%%%%
The latter characteristic feature significantly distinguishes our \textsl{LL-brane}
models from the previously proposed {\em tensionless} $p$-branes (for a review,
see Ref.\ct{lindstroem-etal}). The latter rather resemble $p$-dimensional continuous
distributions of independent massless point-particles without cohesion among the latter.

(c) Consistency of \textsl{LL-brane} dynamics in a spherically or axially
symmetric gravitational background of codimension one requires the presence
of an event horizon which is automatically occupied by the \textsl{LL-brane}
(``horizon straddling'' according to the terminology of Ref.\ct{Barrabes-Israel}).

(d) When the \textsl{LL-brane} moves as a {\em test} brane in spherically or 
axially symmetric gravitational backgrounds its dynamical tension exhibits 
exponential ``inflation/deflation'' time behavior \ct{inflation-all}
-- an effect similar to the ``mass inflation'' effect around black hole horizons
\ct{israel-poisson}. 

An intriguing novel application of \textsl{LL-branes} as natural self-consistent 
gravitational sources for {\em wormhole} space-times has been developed  in a series of 
recent papers \ct{our-WH,rot-WH,ER-bridge,varna-09}.
In what follows, when discussing wormholes we will have in mind precisely
this physically important class of ``thin-shell'' traversable Lorentzian wormholes 
first introduced by Visser \ct{visser-thin,visser-book}. For a comprehensive general
review of wormhole space-times, we refer to \ct{visser-book,WH-rev}.

In the present work we describe a concise systematic scheme for constructing solutions
of Einstein-Maxwell-Kalb-Ramond gravity-matter system in bulk space-time coupled  
self-consistently to one or more (widely separated) codimension-one electrically 
charged \textsl{LL-branes}. The solutions describe bulk space-time
manifolds consisting of several space-time regions (``universes'') with different
(in general) geometries such that: (i) each separate ``universe'' is a ``vacuum''
solution of Einstein-Maxwell-Kalb-Ramond equations (\textsl{i.e.}, without the 
presence of \textsl{LL-branes}); (ii) the separate ``universes'' are pairwise matched 
(glued together) along some of their common horizons; (iii) each of these
common matching horizons is automatically occupied by one \textsl{LL-brane}
(``horizon straddling'') which generates space-time varying cosmological
constants in the various matching ``universes''.

We present several explicit types of solutions with different physical
interpretation such as: (a) wormhole-like space-times with one, two or more
``throats''; (b) non-singular black holes; (c) brane worlds; (d) space-times
undergoing a sequence of spontaneous compactification/decompactification
transitions triggered by \textsl{LL-branes}.

%%%%%%%%%%%%%%%%%%%%%%%%%%%%%%%%%%%%%%%%%%%%%%%%%%%%%%%%%%%%%
%%%%%%%%%%%%%%%%%%%%%%%%%%%%%%%%%%%%%%%%%%%%%%%%%%%%%%%%%%%%%
\section{Lagrangian Formulation of Lightlike Brane Dynamics}
In a series of previous papers
\ct{LL-brane-main,inflation-all,our-WH,rot-WH,ER-bridge,reg-BH}
we have proposed manifestly reparametrization invariant world-volume Lagrangian 
formulation of \textsl{LL-branes} in several dynamically equivalent forms.
Here we will use the Nambu-Goto-type formulation given by the world-volume action:
\be
S_{\rm LL} = - \int d^{p+1}\s \, T 
\sqrt{\bgv\, \det\Vert g_{ab} - \eps \frac{1}{T^2}\pa_a u \pa_b u\Vert\,\bgv}
\quad ,\quad \eps = \pm 1 \; .
\lab{LL-action}
\ee
Here and below the following notations are used:
\begin{itemize}
\item
$g_{ab}$ is the induced metric on the world-volume:
\be
g_{ab} \equiv \pa_a X^{\m} \pa_b X^{\n} G_{\m\n}(X) \; ,
\lab{ind-metric}
\ee
which becomes {\em singular} on-shell (manifestation of the lightlike nature, 
cf. Eq.\rf{on-shell-singular-A} below).
\item
$X^\m (\s)$ are the $p$-brane embedding coordinates in the bulk
$D$-dimensional space-time with Riemannian metric
$G_{\m\n}(X)$ ($\m,\n = 0,1,\ldots ,D-1$); 
$(\s)\equiv \(\s^0 \equiv \t,\s^i\)$ with $i=1,\ldots ,p$;
$\pa_a \equiv \partder{}{\s^a}$.
\item
$u$ is auxiliary world-volume scalar field defining the lightlike direction
(see Eq.\rf{on-shell-singular-A} below); the choice of the sign of $\eps$ in 
\rf{LL-action} does not have physical effect because of the non-propagating 
nature of the $u$-field (see Appendix).
\item
$T$ is {\em dynamical (variable)} brane tension (also a non-propagating
degree of freedom, cf. Appendix).
\end{itemize}

The corresponding equations of motion w.r.t. $X^\m$, $u$ and $T$ read accordingly
(with $\G^\m_{\l\n}$ -- Christoffel connection for the bulk metric):
\br
\pa_a \( T \sqrt{|{\wti g}|} {\wti g}^{ab}\pa_b X^\m\)
+ T \sqrt{|{\wti g}|} {\wti g}^{ab} \pa_a X^\l \pa_b X^\n \G^\m_{\l\n} = 0 \; ,
\lab{X-eqs-NG}\\
% \pa_a \(\frac{1}{T} \sqrt{|{\wti g}|} {\wti g}^{ab}\pa_b u\) = 0 \; ,
% \lab{u-eqs-NG} \\
% T^2 + \eps {\wti g}^{ab}\pa_a u \pa_b u = 0 \; ,
% \lab{T-eq-NG}
\pa_a \(\frac{1}{T} \sqrt{|{\wti g}|} {\wti g}^{ab}\pa_b u\) = 0 \quad ,\quad
T^2 + \eps {\wti g}^{ab}\pa_a u \pa_b u = 0 \; ,
\lab{u-T-eqs-NG}
\er
where we have introduced the convenient notations:
\be
{\wti g}_{ab} = g_{ab} - \eps \frac{1}{T^2}\pa_a u \pa_b u \quad ,\quad
{\wti g} \equiv \det\Vert {\wti g}_{ab}\Vert \; ,
\lab{ind-metric-ext}
\ee
and ${\wti g}^{ab}$ is the inverse matrix w.r.t. ${\wti g}_{ab}$.

From the definition \rf{ind-metric-ext} and second Eq.\rf{u-T-eqs-NG} one easily finds
that the induced metric on the world-volume is singular on-shell:
\be
g_{ab} \( {\wti g}^{bc}\pa_c u\) = 0
\lab{on-shell-singular-A}
\ee
exhibiting the lightlike nature of the $p$-brane described by \rf{LL-action}.

Similarly to the ordinary bosonic $p$-brane we can
rewrite the Nambu-Goto-type action for the \textsl{LL-brane} \rf{LL-action} in a
Polyakov-like form by employing an {\em intrinsic} Riemannian world-volume metric
$\g_{ab}$:
\be
S_{\rm LL-Pol} = - \h \int d^{p+1}\s\, T b_0^{\frac{p-1}{2}}\sqrt{-\g}
\llb \g^{ab} \( g_{ab} - \eps\frac{1}{T^2} \pa_a u \pa_b u\) - \eps b_0 (p-1)\rrb \; ,
\lab{LL-action-dual}
\ee
where $b_0$ is a positive constant. % (for the meaning of $b_0$ see \rf{} below). 
The world-volume action \rf{LL-action-dual} produces the same equations of motion 
\rf{X-eqs-NG}--\rf{u-T-eqs-NG} together with the relation:
\be
\g_{ab} = \frac{\eps}{b_0} {\wti g}_{ab} \; .
\lab{Pol-NG-rel}
\ee
In particular, relation \rf{Pol-NG-rel} reveals the meaning of $b_0$ as
(inverse) proportionality factor between the intrinsic world-volume metric and the
``extended'' induced metric \rf{ind-metric-ext}.

\textbf{Remark.} Let us note that consistency between the Lorentz nature of
the intrinsic world-volume metric $\g_{ab}$ and the Lorentz nature of the
embedding space-time metric $G_{\m\n}$, taking into account \rf{Pol-NG-rel}, requires
to set $\eps = 1$ in the Polyakov-type action \rf{LL-action-dual}.

As shown in our previous papers \ct{LL-brane-main,inflation-all,our-WH}, using the 
above world-volume Lagrangian framework one can add in a natural way 
couplings of the \textsl{LL-brane} to bulk space-time Maxwell $\cA_\m$ 
and Kalb-Ramond $\cA_{\m_1\ldots\m_{D-1}}$ gauge fields (the latter -- in the case of 
codimension one \textsl{LL-branes}, \textsl{i.e.}, for $D=(p+1)+1$). For the
Nambu-Goto-type action \rf{LL-action} these couplings read 
(second ref.\ct{BR-WH-kink}):
\br
{\wti S}_{\rm LL} \lb q,\b\rb = - \int d^{p+1}\s \, T 
\sqrt{\bgv\, \det\Vert g_{ab} - \frac{1}{T^2}
(\pa_a u + q\cA_a)(\pa_b u  + q\cA_b)\Vert\,\bgv}
\nonu \\
- \frac{\b}{(p+1)!} \int d^{p+1}\s\,\vareps^{a_1\ldots a_{p+1}}
\pa_{a_1} X^{\m_1}\ldots\pa_{a_{p+1}} X^{\m_{p+1}} \cA_{\m_1\ldots\m_{p+1}}
\lab{LL-action+EM+KR}
\er
with $g_{ab}$ denoting the induced metric on the world-volume \rf{ind-metric}
and $\cA_a \equiv \pa_a X^\m \cA_\m$.
Using the short-hand notation generalizing \rf{ind-metric-ext}:
\be
{\bar g}_{ab} \equiv g_{ab} - \eps \frac{1}{T^2}
(\pa_a u + q\cA_a)(\pa_b u  + q\cA_b) 
\quad , \quad \cA_a \equiv \pa_a X^\m \cA_\m \; ,
\lab{ind-metric-ext-A}
\ee
% with ${\wti g}^{ab}$ indicating its inverse, 
the equations of motion w.r.t. $X^\m$, $u$ and $T$ acquire the form:
\br
\pa_a \( T \sqrt{|{\bar g}|} {\bar g}^{ab}\pa_b X^\m\)
+ T \sqrt{|{\bar g}|} {\bar g}^{ab} \pa_a X^\l \pa_b X^\n \G^\m_{\l\n}
\nonu \\
+ \eps \frac{q}{T} \sqrt{|{\bar g}|} {\bar g}^{ab}
\pa_a X^\n (\pa_b u  + q\cA_b) \cF^{\l\n}G^{\m\l}
\nonu \\
- \frac{\b}{(p+1)!} \vareps^{a_1\ldots a_{p+1}} \pa_{a_1} X^{\m_1} \ldots
\pa_{a_{p+1}} X^{\m_{p+1}} \cF_{\l\m_1\dots\m_{p+1}} G^{\l\m} = 0 \; ,
\lab{X-eqs-NG-A}
\er
with
\be
\cF_{\m\n} = \pa_\m \cA_\n - \pa_\n \cA_\m \quad ,\quad
\cF_{\m_1\ldots\m_D} = D\pa_{[\m_1} \cA_{\m_2\ldots\m_D]} =
\cF \sqrt{-G} \vareps_{\m_1\ldots\m_D}
\lab{F-KR}
\ee
being the field-strengths of the electromagnetic $\cA_\m$ and Kalb-Ramond \\ 
$\cA_{\m_1\ldots\m_{D-1}}$ gauge potentials \ct{aurilia-townsend}, and
\br
\pa_a \(\frac{1}{T} \sqrt{|{\bar g}|} {\bar g}^{ab}(\pa_b u  + q\cA_b)\) = 0
% \; ,
% \lab{u-eqs-NG-A} \\
\;\; ,\;\;
T^2 + \eps {\bar g}^{ab}(\pa_a u  + q\cA_a)(\pa_b u  + q\cA_b) = 0 \; .
% \lab{T-eq-NG-A}
\nonu \\
{}
\lab{u-T-eqs-NG-A}
\er
The on-shell singularity of the induced metric $g_{ab}$ \rf{ind-metric}, 
\textsl{i.e.}, the lightlike property, now reads (using notation 
\rf{ind-metric-ext-A}, cf. Eq.\rf{on-shell-singular-A}):
\be
g_{ab} \({\bar g}^{bc}(\pa_c u  + q\cA_c)\) = 0 \; .
\lab{on-shell-singular-A-A}
\ee

The Polyakov-type form of the world-volume action \rf{LL-action+EM+KR} becomes
(using short-hand notation \rf{ind-metric-ext-A}):
\br
{\wti S}_{\rm LL-Pol} \lb q,\b\rb =
- \h \int d^{p+1}\s\, T b_0^{\frac{p-1}{2}}\sqrt{-\g}
\llb \g^{ab} {\bar g}_{ab} - \eps b_0 (p-1)\rrb 
\nonu \\
- \frac{\b}{(p+1)!} \int d^{p+1}\s\,\vareps^{a_1\ldots a_{p+1}}
\pa_{a_1} X^{\m_1}\ldots\pa_{a_{p+1}} X^{\m_{p+1}} \cA_{\m_1\ldots\m_{p+1}} \; , 
\lab{LL-action+EM+KR-dual}
\er
yielding the same set of equations of motion \rf{X-eqs-NG-A}--\rf{u-T-eqs-NG-A} plus the
counterpart of \rf{Pol-NG-rel}:
\be
\g_{ab} = \frac{\eps}{b_0} {\bar g}_{ab} 
\lab{Pol-NG-rel+EM+KR}
\ee
with ${\bar g}_{ab}$ as in \rf{ind-metric-ext-A}. Here again the above
remark after Eq.\rf{Pol-NG-rel} applies, \textsl{i.e.}, that for consistency we must
set $\eps = 1$ within the Polyakov-type action \rf{LL-action+EM+KR-dual}.

%%%%%%%%%%%%%%%%%%%%%%%%%%%%%%%%%%%%%%%%%%%%%%%%%%%%%%%%%%%%%
%%%%%%%%%%%%%%%%%%%%%%%%%%%%%%%%%%%%%%%%%%%%%%%%%%%%%%%%%%%%%
\section{Bulk Gravity/Gauge-Field System Self-Consistently Interacting
With Lightlike Branes}
\subsection{Lagrangian Formulation}
Let us now consider self-consistent bulk Einstein-Maxwell-Kalb-Ramond system coupled 
to $N\geq 1$ distantly separated charged codimension-one {\em lightlike}
$p$-branes (in this case $D=(p+1)+1$). The pertinent Lagrangian action reads:
\br
S = \int\!\! d^D x\,\sqrt{-G}\,\llb \frac{R(G)}{16\pi} 
- \frac{1}{4} \cF_{\m\n}\cF^{\m\n} 
- \frac{1}{D! 2} \cF_{\m_1\ldots\m_D}\cF^{\m_1\ldots\m_D}\rrb 
\nonu \\
+ \sum_{k=1}^N {\wti S}_{\mathrm{LL}}\lb q^{(k)},\b^{(k)} \rb \; ,
\lab{E-M-KR+2LL}
\er
where again $\cF_{\m\n}$ and $\cF_{\m_1\ldots\m_D}$ are the Maxwell and Kalb-Ramond
field-strengths \rf{F-KR} and 
${\wti S}_{\mathrm{LL}}\lb q^{(k)},\b^{(k)}\rb$ indicates the world-volume action
of the $k$-th \textsl{LL-brane} of the form \rf{LL-action+EM+KR}
(or \rf{LL-action+EM+KR-dual}).

The corresponding equations of motion are as follows:

\vspace{.1in}
(a) Einstein equations:
\be
R_{\m\n} - \h G_{\m\n} R =
8\pi \( T^{(EM)}_{\m\n} + T^{(KR)}_{\m\n} + 
\sum_{k=1}^N T^{(brane-k)}_{\m\n}\) \; .
\lab{Einstein-eqs}
\ee
The energy-momentum tensors of bulk gauge fields are given by:
\be
T^{(EM)}_{\m\n} = \cF_{\m\k}\cF^{\m\n} - G_{\m\n}\frac{1}{4}\cF_{\k\l}\cF^{\k\l} 
\quad ,\quad T^{(KR)}_{\m\n} = - \h \cF^2 G_{\m\n} \; , 
\lab{T-EM-KR}
\ee
where the last relation indicates that $\L \equiv 4\pi \cF^2$ can be
interpreted as dynamically generated cosmological ``constant''.
The energy-momentum (stress-energy) tensor of $k$-th \textsl{LL-brane} is
straightforwardly derived from the pertinent \textsl{LL-brane} action 
\rf{LL-action+EM+KR}:
\be
T_{(brane-k)}^{\m\n} = 
- \int\!\! d^{p+1}\s\,\frac{\d^{(D)}\Bigl(x-X_{(k)}(\s)\Bigr)}{\sqrt{-G}}
\, T^{(k)}\,\sqrt{|{\bar g}_{(k)}|} {\bar g}_{(k)}^{ab}
\pa_a X_{(k)}^\m \pa_b X_{(k)}^\n \; ,
\lab{T-brane-A}
\ee
where for each $k$-th \textsl{LL-brane}:
\br
{\bar g}^{(k)}_{ab} \equiv g^{(k)}_{ab} - \eps^{(k)} \frac{1}{T_{(k)}^2}
(\pa_a u^{(k)} + q^{(k)}\cA^{(k)}_a)(\pa_b u^{(k)}  + q^{(k)}\cA^{(k)}_b) 
\nonu \\
g^{(k)}_{ab} = \pa_a X_{(k)}^\m G_{\m\n} \pa_b X_{(k)}^\n \quad , \quad 
\eps^{(k)} = \pm 1 \quad , \quad \cA^{(k)}_a \equiv \pa_a X_{(k)}^\m \cA_\m \; .
\lab{ind-metric-ext-A-k}
\er

\vspace{.1in}
(b) Maxwell equations:
\br
\pa_\n \(\sqrt{-G}\cF^{\m\n}\) -
\sum_{k=1}^N q^{(k)} \int\!\! d^{p+1}\s\,\d^{(D)}\Bigl(x-X_{(k)}(\s)\Bigr)
\nonu \\
\times \sqrt{|{\bar g}_{(k)}|} {\bar g}_{(k)}^{ab}\pa_a X_{(k)}^\m 
\frac{\pa_b u^{(k)} + q^{(k)}\cA^{(k)}_b}{T^{(k)}}
= 0 \; ,
\lab{Maxwell-eqs}
\er
using notations \rf{ind-metric-ext-A-k}.

\vspace{.1in}
(c) Kalb-Ramond equations of motion (recall definition of $\cF$ in \rf{F-KR}):
% $\cF_{\m_1\ldots\m_D} = % D\pa_{[\m_1} \cA_{\m_2\ldots\m_D]} =
% \cF \sqrt{-G} \vareps_{\m_1\ldots\m_D}$):
\br
\vareps^{\n\m_1\ldots\m_{p+1}} \pa_\n \cF
- \sum_{k=1}^N \b^{(k)}\,\int\! d^{p+1}\s\,\d^{(D)}(x - X_{(k)}(\s))
\nonu \\
\times \vareps^{a_1\ldots a_{p+1}} \pa_{a_1} X_{(k)}^{\m_1}\ldots
\pa_{a_{p+1}} X_{(k)}^{\m_{p+1}} = 0 \; .
\lab{F-KR-eqs}
\er

\vspace{.1in}
(d) The \textsl{LL-brane} equations of motion have already been written down in
\rf{X-eqs-NG-A}--\rf{u-T-eqs-NG-A} above.

\subsection{LL-Brane Dynamics in Static ``Spherically Symmetric'' Backgrounds}
We will be interested in static ``spherically-symmetric''-type solutions of 
Einstein-Maxwell-Kalb-Ramond equations with the following generic form of
the bulk Riemannian metric: 
\be
ds^2 = - A(\eta) dt^2 + \frac{d\eta^2}{A(\eta)} + 
C(\eta) h_{ij}(\th) d\th^i d\th^j \; ,
\lab{static-spherical}
\ee
or, in Eddington-Finkelstein coordinates ($dt = dv-\frac{d\eta}{A(\eta)}$) :
\be
ds^2 = - A(\eta) dv^2 + 2 dv\,d\eta + C(\eta) h_{ij}(\th) d\th^i d\th^j \; .
\lab{EF-metric}
\ee
Here $h_{ij}$ indicates the standard metric on $p$-dimensional sphere, cylinder,
torus or flat Euclidean section.
The ``radial-like'' coordinate $\eta$ will vary in general from $-\infty$ to $+\infty$. 

We will consider the simplest ansatz for the \textsl{LL-brane} embedding
coordinates:
\be
X^0\equiv v = \t \quad, \quad X^1\equiv \eta = \eta (\t) \quad, \quad 
X^i\equiv \th^i = \s^i \;\; (i=1,\ldots ,p) \; .
\lab{X-embed}
\ee
Furthermore, we will use explicit world-volume reparametrization invariance
of the \textsl{LL-brane} actions (\rf{LL-action-dual} and \rf{LL-action+EM+KR-dual})
to introduce the standard synchronous gauge-fixing conditions for the intrinsic
world-volume metric:
\be
\g^{00} = -1 \quad ,\quad \g^{0i} = 0 \;\; (i=1,\ldots,p) \; .
\lab{gauge-fix}
\ee
The latter together with second Eq.\rf{u-T-eqs-NG-A} and \rf{Pol-NG-rel+EM+KR} 
(and accounting for the definition \rf{ind-metric-ext-A}) implies for the 
$00$-component of the induced metric \rf{ind-metric} on the \textsl{LL-brane} 
world-volume:
\be
g_{00} \equiv \Xdot^\m\!\! G_{\m\n}\!\!\Xdot^\n = 
\frac{b_0}{T^2} {\bar g}^{ij} \(\pa_i u + \cA_i\)\(\pa_j u + \cA_j\) \geq 0
\lab{g-00}
\ee
which must match the condition $g_{00} \leq 0$ required by consistency
between the Lorentz form of the bulk space-time metric and the Lorentz form
of the \textsl{LL-brane} world-volume metric. Hence we are led to impose the
ansatz:
\be
\pa_i u + \cA_i = 0
\lab{u-ansatz}
\ee
which is consistent for static spherically symmetric bulk space-time Maxwell field 
$\cA_\m$ and whose physical meaning is that the lightlike direction for the induced 
metric in Eq.\rf{on-shell-singular-A-A} (or Eq.\rf{on-shell-singular-A} for
electrically neutral \textsl{LL-brane}) coincides with the brane proper-time 
$\t$-direction on the world-volume.

Thus, taking into account \rf{gauge-fix} and \rf{u-ansatz}, the \textsl{LL-brane} 
equations of motion \rf{u-T-eqs-NG-A} (or, equivalently, \rf{on-shell-singular-A-A})
reduce to:
\br
g_{00} \equiv \Xdot^\m\!\! G_{\m\n}\!\! \Xdot^\n = 0 
% \quad ,\quad g_{0i} = 0 
\quad ,\quad g_{0i} \equiv \Xdot^\m\!\! G_{\m\n} \pa_i X^\n = 0 \; ,
\lab{g-eqs-A} \\
T^2 = \frac{1}{b_0} \(\pa_0 u + \cA_0\)^2 % \;\; (\;\mathrm{i.e.}\;
\; ,\; \pa_i T = 0 % \; ,
% \lab{T-eqs-A} \\
\;\; ,\;\;
\pa_0 g^{(p)} = 0 \;\; \( g^{(p)}\equiv\det\Vert g_{ij}\Vert\) \; ,
\lab{g-p-0}
\er
with $g_{ij}$ being the spacelike part of the induced metric \rf{ind-metric}.
Eqs.\rf{g-eqs-A}--\rf{g-p-0} with \textsl{LL-brane} embedding \rf{X-embed}
and metric of the form \rf{EF-metric} imply: 
\be
-A(\eta) + 2\etadot = 0 \quad , \quad 
\pa_\t C = \etadot\,\pa_\eta C\bv_{\eta=\eta(\t)} = 0 \; .
\lab{eta-const}
\ee
Here we will distinguish two cases.
First, let us consider the case of $C(\eta)$ as non-trivial function of $\eta$ 
(\textsl{i.e.}, proper spherically-symmetric-type space-time). In this case 
Eqs.\rf{eta-const} imply:
\be
\etadot = 0 \; \to \; \eta (\t) = \eta_0 = \mathrm{const} \quad ,\quad 
A(\eta_0) = 0 \; .
\lab{horizon-standard}
\ee
Eq.\rf{horizon-standard} tells us that consistency of \textsl{LL-brane} dynamics
in a proper spherically-symmetric-type gravitational background of codimension
one requires the latter to possess a horizon (at some $\eta = \eta_0$), which 
is automatically occupied by the \textsl{LL-brane} (``horizon straddling'' 
according to the terminology of Ref.\ct{Barrabes-Israel}). Similar property -- 
``horizon straddling'', has been found also for \textsl{LL-branes} moving in
rotating axially symmetric (Kerr or Kerr-Newman) and rotating cylindrically
symmetric black hole backgrounds \ct{our-WH,rot-WH}. 

Next, consider the case $C(\eta) = \mathrm{const}$ in \rf{EF-metric},
\textsl{i.e.}, the
corresponding space-time manifold is of product type $\S_2 \times S^p$.
A physically relevant example is the Bertotti-Robinson \ct{BR,lapedes-78}
space-time in $D=4$ (\textsl{i.e.}, $p=2$) describing 
Anti-de-Sitter${}_2 \times S^2$ with metric (in Eddington-Finkelstein
coordinates):
\be
ds^2 = -\frac{\eta^2}{r_0^2}dv^2 + 2 dv d\eta + % r_0^2 h_{ij}(\th) d\th^i d\th^j \, .
r_0^2 \llb d\th^2 + \sin^2 \th d\vp^2 \rrb \; .
\lab{BR-EF}
\ee
At $\eta = 0$ the Bertotti-Robinson metric \rf{BR-EF} possesses a 
horizon. Further, we will consider the case of Bertotti-Robinson universe with
constant electric field $\cF_{v\eta} = \pm \frac{1}{2 r_0 \sqrt{\pi}}$.
% \be
% \cF_{v\eta} = \pm \frac{1}{2 r_0 \sqrt{\pi}} \; .
% \lab{BR-Maxwell}
% \ee
In the present case the second Eq.\rf{eta-const} is trivially satisfied whereas the
first one yields:
$\eta (\t) = \eta (0) \Bigl(1 - \t\frac{\eta (0)}{2 r_0^2}\Bigr)^{-1}$.
% \be
% \eta (\t) = \frac{\eta (0)}{1 - \t\frac{\eta (0)}{2 r_0^2}} \; .
% \lab{BR-eta}
% \ee
In particular, if the \textsl{LL-brane} is initially (at $\t=0$) located on the 
Bertotti-Robinson horizon $\eta = 0$, it will stay there permanently. It is
this particular solution which we will consider in what follows.

%%%%%%%%%%%%%%%%%%%%%%%%%%%%%%%%%%%%%%%%%%%%%%%%%%%%%%%%%%%%%
%%%%%%%%%%%%%%%%%%%%%%%%%%%%%%%%%%%%%%%%%%%%%%%%%%%%%%%%%%%%%
\section{Self-Consistent Wormhole-Like Solutions with LL-Branes -- General Scheme}
We will construct self-consistent static ``spherically symmetric''  solutions of 
the system of Einstein-Maxwell-Kalb-Ramond equations \rf{Einstein-eqs}--\rf{F-KR-eqs}
and \textsl{LL-brane} Eqs.\rf{X-eqs-NG-A}--\rf{u-T-eqs-NG-A}
% using 
% the \textsl{LL-brane} embedding \rf{} and the gauge-fixing conditions
% of the \textsl{LL-brane} world-volume reparametrization invariance \rf{}
following the steps:

\vspace{.1in}
(i) The bulk space-time metric will be of the form:
\br
ds^2 = - A(\eta) dv^2 + 2 dv\,d\eta + C(\eta) h_{ij}(\th) d\th^i d\th^j \; ,
\nonu \\
A(\eta^{(k)}_0) = 0 \;\; (k=1,\ldots ,N)\quad ,\quad 
A(\eta) > 0 \;\; \mathrm{for ~all}\; \eta \neq \eta^{(k)}_0
\lab{eqsys-0}
\er
Each horizon at $\eta = \eta^{(k)}_0$ is automatically occupied by (one of
the) \textsl{LL-brane(s)} according to the \textsl{LL-brane} dynamics 
(``horizon straddling'', cf.\rf{eta-const}--\rf{horizon-standard}).

\vspace{.1in}
(ii) Choose ``vacuum'' solutions of Einstein-Maxwell-Kalb-Ramond
equations \rf{Einstein-eqs}--\rf{F-KR-eqs}
(\textsl{i.e.}, without the delta-function terms due to the
\textsl{LL-branes}) in each region $-\infty <\eta<\eta^{(1)}_0\, ,\,
\eta^{(1)}_0 <\eta < \eta^{(2)}_0 ,\ldots, \eta^{(N)}_0 <\eta <\infty$.

\vspace{.1in}
(iii) Match the discontinuities across each horizon at $\eta = \eta^{(k)}_0$ 
of the derivatives of the bulk metric, Maxwell and Kalb-Ramond field strengths
using the explicit expressions for the \textsl{LL-brane} stress-energy tensors, 
electric and Kalb-Ramond currents systematically derived from the underlying 
\textsl{LL-brane} world-volume actions \rf{LL-action+EM+KR-dual}.

In particular, for the stress-energy tensor of each $k$-th \textsl{LL-brane} we
obtain (here we suppress the index $(k)$):
\be
T_{(brane)}^{\m\n} = S^{\m\n}\,\d (\eta-\eta_0)
\lab{T-S-0}
\ee
with surface energy-momentum tensor:
\be
S^{\m\n} \equiv \frac{T}{\eps b_0^{1/2}}\,
\( \pa_\t X^\m \pa_\t X^\n - \eps b_0 G^{ij} \pa_i X^\m \pa_j X^\n 
\)_{v=\t,\,\eta=\eta_0,\,\th^i =\s^i} \; ,
\lab{T-S-brane} 
\ee
where $G_{ij} = C(\eta) h_{ij}(\th)$ (cf. \rf{EF-metric}). For the non-zero 
components of \rf{T-S-brane} (with lower indices) and its trace we find:
\be
S_{\eta\eta} = \eps \frac{T}{b_0^{1/2}} \quad ,\quad 
S_{ij} = - T b_0^{1/2} G_{ij} \quad ,\quad S^\l_\l = - pTb_0^{1/2} \; .
\lab{S-comp}
\ee

\vspace{.1in}
Taking into account \rf{T-S-0}--\rf{S-comp} Einstein equations \rf{Einstein-eqs}
yield:
\be
\llb \pa_\eta A \rrb_{\eta^{(k)}_0} = - 16\pi T^{(k)} \sqrt{b^{(k)}_0} 
\quad,\quad 
\llb \pa_\eta \ln C \rrb_{\eta^{(k)}_0} = 
- \frac{16\pi}{p\sqrt{b^{(k)}_0}} T^{(k)}
\lab{eqsys-1-2}
\ee
with notation $\bigl\lb Y \bigr\rb_{\eta_0} \equiv 
Y\bv_{\eta \to \eta_0 +0} - Y\bv_{\eta \to \eta_0 -0}$ for any quantity $Y$.

Maxwell and Kalb-Ramond equations yield:
\be
\llb \cF_{v\eta} \rrb_{\eta^{(k)}_0} = q^{(k)} \quad ,\quad
\llb \cF \rrb_{\eta^{(k)}_0} = - \b^{(k)}
\lab{eqsys-4-5}
\ee
In Eqs.\rf{eqsys-1-2}--\rf{eqsys-4-5} $\( T^{(k)},b^{(k)}_0 \)$ indicate the
dynamical tension and $b_0$ parameter of the 
$k$-th \textsl{LL-brane} occupying horizon $\eta^{(k)}_0$, with electric charge
surface density $q^{(k)}$ and Kalb-Ramond coupling $\b^{(k)}$. The second
relation in \rf{eqsys-4-5} gives the jump of the dynamically generated
cosmological constant $\L \equiv 4\pi \cF^2$ across the $k$-th \textsl{LL-brane}.

The only non-trivial contribution of \textsl{LL-brane} equations of motion
comes from the $X^0$-equation which yields:
\br
\pa_0 T^{(k)} + T^{(k)} \h \( \llangle \pa_\eta A \rrangle_{\eta^{(k)}_0} 
+ p b^{(k)}_0 \llangle \pa_\eta \ln C \rrangle_{\eta^{(k)}_0} \)
\nonu \\
-  \sqrt{b^{(k)}_0} \( q^{(k)} \llangle \cF_{v\eta}\rrangle_{\eta^{(k)}_0} 
- \b^{(k)} \llangle \cF\rrangle_{\eta=^{(k)}_0}\) = 0
\lab{eqsys-3}
\er
with notation $\llangle Y \rrangle_{\eta_0} \equiv 
\h \( Y\bv_{\eta \to \eta_0 +0} + Y\bv_{\eta \to \eta_0 -0}\)$. 
% In \rf{eqsys-3} we have used the prescrition

In what follows we will take time-independent dynamical \textsl{LL-brane}
tension(s) ($\pa_0 T^{(k)}=0$) because of matching static bulk space-time 
geometries. 
%%%%%%%%%%
Let us also note that the appearance of mean values of the
corresponding quantities with discontinuities across the horizons follows
the resolution of the discontinuity problem given in \ct{Israel-66} (see also
\ct{BGG}).

The wormhole-like solutions presented in the next Section share the
following important properties:

(a) The \textsl{LL-branes} at the wormhole ``throats'' represent ``exotic'' 
matter -- $T\leq 0$, \textsl{i.e.}, negative or zero brane tension implying 
violation of null-energy conditions as predicted by general wormhole arguments
\ct{visser-book} (although the latter could be remedied via quantum fluctuations).

(b) The wormhole-like space-times constructed via \textsl{LL-branes} at
their ``throats'' are {\em not} traversable w.r.t.
the ``laboratory'' time of a static observer in either of the different
``universes'' comprising the pertinent wormhole space-time manifold.
On the other hand, they {\em are traversable} w.r.t. the {\em proper time} 
of a traveling observer.

Proper-time traversability can be easily seen by considering dynamics of 
test particle of mass $m_0$ (``traveling observer'') in a wormhole background, 
which is described by the world-line action:
\be
S_{\mathrm{particle}} = \h \int d\l \Bigl\lb\frac{1}{e}\xdot^\m \xdot^\n G_{\m\n}
- e m_0^2 \rb \; .
% S_{\mathrm{particle}} = \int d\l \Bigl\lb\h\Bigl(\frac{1}{e}\xdot^\m \xdot^\n G_{\m\n}
% - e m_0^2\Bigr) - q_0 \xdot^\m \cA_\m\Bigr\rb \; .
\lab{test-particle}
\ee
Using energy $\cE$ and orbital momentum $\cJ$ conservation and introducing the 
{\em proper} world-line time $s$ ($\frac{ds}{d\l}= e m_0$), the ``mass-shell''
equation (the equation w.r.t. the ``einbein'' $e$ produced by the action 
\rf{test-particle}) yields:
\be
% \eta^{\pr\, 2} + \cV_{\mathrm{eff}} (\eta) = \frac{\cE^2}{m_0^2}
\(\frac{d\eta}{ds}\)^2 + \cV_{\mathrm{eff}} (\eta) = \frac{\cE^2}{m_0^2}
\quad ,\quad 
\cV_{\mathrm{eff}} (\eta) \equiv A(\eta) \Bigl( 1 + \frac{\cJ^2}{m_0^2 C(\eta)}\Bigr) 
\lab{particle-eq-2}
\ee
where the metric coefficients $\cA (\eta),\, C(\eta)$ are those in
\rf{eqsys-0}.
Irrespectively of the specific form of the ``effective potential'' in
\rf{particle-eq-2}, a ``radially'' moving (with zero ``impact'' parameter
$\cJ=0$) traveling observer (and with sufficiently large energy $\cE$) will
always cross within finite amount of proper-time through any ``throat'' 
$(\eta = \eta_0^{(k)})$ from one ``universe'' to another and possibly even
shuttle between them (cf. Subsection 5.4 below).

%%%%%%%%%%%%%%%%%%%%%%%%%%%%%%%%%%%%%%%%%%%%%%%%%%%%%%%%%%%%%
%%%%%%%%%%%%%%%%%%%%%%%%%%%%%%%%%%%%%%%%%%%%%%%%%%%%%%%%%%%%%
\section{Examples}
Henceforth we will use the following acronyms for brevity:
``BR''=``Bertotti-Robinson'', ``Schw''=``Schwarzschild'', 
``RN''= ``Reissner-Nordstr{\"o}m'', \\``(A)dS''=``(Anti-)de-Sitter'',
``SdS'' = ``Schwarzschild-de-Sitter'', and \textsl{LL-brane} matching will
be denoted by ``$|$''.
%%%%%%%%%%%%%%%%%%%%%%%%%%%%%%%%%%%%%%%%%%%%%%%%%%%%%%%%%%%%%
\subsection{Symmetric Wormhole with Reissner-Nordstr{\"o}m Geometry}
It consists of two identical copies of exterior % Reissner-Nordstr{\"o}m 
RN region ($r > r_0$, 
$r_0$ denoting the {\em outer} RN horizon) -- ``left'' RN ``universe'' 
($\eta<0$) and ``right'' RN ``universe'' ($\eta>0$) glued together via a
LL-brane sitting on $r=r_0$ ($\eta=0$):
\br
ds^2 = - A(\eta) dv^2 + 2 dv\,d\eta + 
C(\eta) \llb d\th^2 + \sin^2 \th\, d\varphi^2 \rrb \; ,
\lab{RN-RN-1} \\
A(\eta) = 1 - \frac{2m}{r_0 + |\eta|} + \frac{Q^2}{(r_0 + |\eta|)^2} \quad, \quad
C(\eta) = (r_0 + |\eta|)^2 \; ,
\lab{RN-RN-2} \\
A(0) = 0 \quad ,\quad A(\eta) > 0 \;\; \mathrm{for} \; \eta \neq 0 \; .
\er
RN mass is determined by the dynamical LL-brane tension $T$:
\be
\bigl( 16\pi\,|T|\sqrt{b_0}\, m - 1\bigr) \( m^2 - Q^2\) + 
16\pi^2 T^2 b_0 Q^4 = 0 \; .
\lab{M-RN}
\ee
In the particular case of Schwarzschild wormhole (Einstein-Rosen ``bridge'', $Q=0$):
$m= 1/8\pi |T|$.

%%%%%%%%%%%%%%%%%%%%%%%%%%%%%%%%%%%%%%%%%%%%%%%%%%%%%%%%%%%%%
\subsection{Non-singular Black Hole}
It is described by the metric:
\br
ds^2 = - A(r) dv^2 + 2 dv\,dr + r^2 \llb d\th^2 + \sin^2 \th\, d\varphi^2 \rrb \; ;
\lab{EF-metric-1} \\
A(r) \equiv A_{(-)}(r) = 1 - K r^2 \quad , \;\; 
\mathrm{for}\; r < r_0 \quad (\mathrm{de ~Sitter}) \; ,
\lab{de-Sitter}\\
A(r) \equiv A_{(+)}(r) = 1 - \frac{2m}{r} + \frac{Q^2}{r^2} 
\quad , \;\; \mathrm{for}\;  r > r_0  \quad (\mathrm{RN})\; ,
\lab{RN-dS}
\er
where $r_0$ is the common horizon $A_{(\pm)}(r_0) = 0\; ,\; r_0 = m - \sqrt{m^2 - Q^2}$
(internal RN).
% \be
% A_{(-)}(r_0) = 0 \quad ,\quad A_{(+)}(r_0) = 0 \quad ,\quad 
% r_0 = m - \sqrt{m^2 - Q^2}\;\; (\mathrm{interna ~RN}) \;.
% \lab{common-horizon}
% \ee

An electrically charged LL-brane occupies the horizon $r=r_0$ and uniquely determines
all parameters %:
% \be
% r_0 = \frac{1}{\sqrt{K}} \quad ,\quad
% m = \frac{2}{\sqrt{K}} \quad ,\quad
% Q^2 = \frac{3}{K} % \quad ,\quad \mathrm{with}\;\; K = \frac{4\pi}{3} \b^2 \; .
% \lab{metric-matching-1}
% \ee
$r_0 = \frac{1}{\sqrt{K}}\, ,\, m = \frac{2}{\sqrt{K}}\, ,\, Q^2 = \frac{3}{K}$,
with $\L = 3K = \frac{4\pi}{3} \b^2$ -- {\em dynamically generated}
cosmological const in the interior de-Sitter region through the Kalb-Ramond
LL-charge $\b$. Apparently there is {\em no} black hole singularity at $r=0$.

% Shape of the ``effective potential'' $A(r)$ -- apparently {\em no} black
% hole singularity at $r=0$:
% \begin{figure}
% \begin{center}
% \includegraphics[width=10cm,keepaspectratio=true]{02_Reg-BH[dS-RN]_1}
% \end{center}
% \end{figure}

%%%%%%%%%%%%%%%%%%%%%%%%%%%%%%%%%%%%%%%%%%%%%%%%%%%%%%%%%%%%%
\subsection{Asymmetric Wormhole -- Schw-dS $|$ RN}
The overall metric is $ds^2 = - A(\eta) dv^2 + 2 dv\,d\eta + 
(r_0 + |\eta|)^2 \llb d\th^2 + \sin^2 \th\, d\varphi^2 \rrb$ with
$A(0) = 0$. Here we have:

(i) ``left universe'' -- exterior region of Schwarzschild-de-Sitter space-time 
above the {\em inner} (Schwarzschild-type) horizon $r_0$:
\be
A(\eta) = 1 - \frac{2m_1}{r_0 - \eta} - K (r_0 - \eta)^2 
\quad \mathrm{for}\; \eta < 0 \; ;
\lab{left-univ}
\ee
(ii) ``right universe'' -- exterior Reissner-Nordstr{\"o}m region beyond the 
{\em outer} RN horizon $r_0$:
\be
A(\eta) = 1 - \frac{2m_2}{r_0 + \eta} + \frac{Q^2}{(r_0 + \eta)^2}
\quad \mathrm{for}\; \eta > 0 \; .
\lab{right-univ}
\ee

Charged LL-brane occupies the common horizon (wormhole ``throat'') and
determines all wormhole parameters via its charges $(q,\b)$:
\br
m_1 = \frac{\sqrt{b_0}}{4\pi |T|}\Bigl( 1 - \frac{b_0 \b^2}{3\pi T^2}\Bigr)
\quad ,\quad
m_2 = \frac{\sqrt{b_0}}{4\pi |T|}\Bigl( 1 + \frac{4 q^2}{\pi T^2}\Bigr) \; ,
\lab{m1-m2} \\
r_0 = \frac{\sqrt{b_0}}{2\pi |T|} \quad, \quad
T^2 = \frac{\b^2 + 4q^2}{2\pi (1-4b_0)} \quad, \quad 
Q^2 = \frac{16\pi}{b_0} q^2 r_0^4 \; .
\lab{T-Q}
\er
including the dynamically generated cosmological const $\L=3K = 4\pi\b^2$ in
the ``left'' universe.

%%%%%%%%%%%%%%%%%%%%%%%%%%%%%%%%%%%%%%%%%%%%%%%%%%%%%%%%%%%%%
\subsection{Compactification/Decompactification Transitions}
These are wormhole-like solution with two widely separated LL-branes sitting
at horizons $\eta = \eta_0 \equiv 0$ and $\eta = {\bar\eta}_0$, with metric:
\br
ds^2 = - A(\eta) dv^2 + 2 dv d\eta + 
C(\eta) \llb d\th^2 + \sin^2 \th d\vp^2 \rrb \, ,
\lab{asymm-WH-EF} \\
A (0) = 0 \; ,\; A ({\bar\eta}_0) = 0 \; ,\; 
{\bar\eta}_0 \equiv {\bar r}_0 - r_0 >0 % \; ,
% \nonu \\
\; ,\;
A(\eta) > 0 \; \mathrm{for}\; \eta \neq 0,\,{\bar\eta}_0 \; ,
\lab{asymm-WH-A}
\er
describing {\em three} pairwise matched space-time regions:
%\begin{itemize}
%\item

(i) ``left'' Bertotti-Robinson ``universe'' ($AdS_2\times S^2$) for $\eta<0$ where:
\be
A(\eta) = \frac{\eta^2}{r_0^2} \quad, \quad C(\eta) = r_0^2 
\quad, \quad \cF_{v\eta} = \pm \frac{1}{2\sqrt{\pi}\,r_0} \; ;
% \quad \mathrm{for}\; \eta < 0 \; ;
\lab{left-BR}
\ee
%\item

(ii) ``middle'' Reissner-Nordstr{\"o}m-de-Sitter ``universe'' for \\
$0 <\eta <{\bar r}_0 - r_0$ with:
\br
A(\eta) = 1 - \frac{2m}{r_0 + \eta} + \frac{Q^2}{(r_0 + \eta)^2} - 
\frac{4\pi\b^2}{3} (r_0 + \eta)^2 \; ,
\lab{mid-RN-0} \\
C(\eta) = (r_0 + \eta)^2 \;\; ,\;\;
\cF_{v\eta} = \frac{Q}{\sqrt{4\pi} (r_0 + \eta)^2} \;\; ,
% \quad \mathrm{for}\; 0 < \eta < {\bar r}_0 - r_0 \;\; ,
\lab{mid-RN-1}
\er
where $r_0$ and ${\bar r}_0$ (${\bar r}_0 > r_0$) are the intermediate (outer 
RN) and the outmost (de-Sitter) horizons of the standard RN-de-Sitter space-time
(note the dynamically generated cosmological const $\L = 4\pi\b^2$
in \rf{mid-RN-0});
%\item

(iii) another ``right''  Bertotti-Robinson ``universe'' ($AdS_2\times S^2$)
for $\eta > {\bar r}_0 - r_0$:
\be
A(\eta) = \frac{\(\eta - {\bar r}_0 + r_0 \)^2}{{\bar r}_0^2} 
\quad, \quad C(\eta) = {\bar r}_0^2 \quad, \quad
\cF_{v\eta} = \pm \frac{1}{2\sqrt{\pi}\,{\bar r}_0} \; .
% \quad \mathrm{for}\; \eta > {\bar r}_0 - r_0 \; .
\lab{right-BR}
\ee
%\end{itemize}

% Shape of ``effective potential'' $A(\eta)$ in two-''throat'' BR-RNdS-BR
% wormhole:
% \begin{figure}
% \begin{center}
% \includegraphics[width=7cm,keepaspectratio=true]{04_BR-SdS-BR[BR-kink]_1}
% \end{center}
% \end{figure}
Traveling observer along $\eta$-direction % with zero impact parameter
will ``shuttle'' between the three ``universes'' crossing consecutively
both LL-branes at the ``throats'' within {\em finite} intervals of
his/her proper time.

%%%%%%%%%%%%%%%%%%%%%%%%%%%%%%%%%%%%%%%%%%%%%%%%%%%%%%%%%%%%%
\subsection{Multi-``throat'' wormhole Schw $|$ SdS $|$ SdS $|$ Schw}
This is a wormhole-like solution with metric:
\br
ds^2 = - A(\eta) dv^2 + 2 dv d\eta + 
(r_0 + \eta)^2 \llb d\th^2 + \sin^2 \th d\vp^2 \rrb
\nonu \\
A(0) = 0 \quad, \quad A\bigl(\pm ({\bar r}_0 - r_0)\bigr) = 0
\nonu
\er
describing {\em four} pairwise matched space-time regions via 3 widely separated 
LL-branes located at $\eta=0$ and $\eta = \pm ({\bar r}_0 - r_0)$:
%\begin{itemize}
%\item

(i) ``left-most'' ($\eta < -({\bar r}_0 - r_0)$) and ``right-most''
($\eta >{\bar r}_0 - r_0$) ``universes'' comprising the exterior
Schwarzschild region beyond the Schwarzschild horizon at ${\bar r}_0$:
\be
A(\eta) = 1 - \frac{{\bar r}_0}{r_0 + |\eta|} \quad \mathrm{for}\;
|\eta| > {\bar r}_0 - r_0  \; ,
\lab{Schw-SdS-SdS-Schw-1}
\ee
%\item

(ii) two ``middle'' ``universes'', for $-({\bar r}_0 - r_0) <\eta <0$ 
and for $0 < \eta < {\bar r}_0 - r_0$ -- two identical copies of the intermediate
region of Schwarzschild-de-Sitter space-time between the inner (Schwarzschild)
horizon at $r_0$ and the outer (de-Sitter) horizon at ${\bar r}_0$:
\br
A(\eta) = 1 - \frac{2m}{r_0 + |\eta|} - \frac{4\pi\b^2}{3} (r_0 + |\eta|)^2
\quad \mathrm{for}\;
|\eta| < {\bar r}_0 - r_0  \; ,
\lab{Schw-SdS-SdS-Schw-2} % \\
% A(0) = 0 \;\; (\mathrm{inner ~SdS ~horizon ~at} ~r_0)
% \nonu \\
% A\bigl(\pm({\bar r}_0 - r_0)\bigr) = 0  
% \;\; (\mathrm{outer ~SdS ~horizon ~at} ~{\bar r}_0) \; ,
% \nonu
\er
where $A(0)=0$ (inner SdS horizon) and $A\bigl(\pm({\bar r}_0 - r_0)\bigr) = 0$
(outer SdS horizon) and 
with dynamically generated (by the LL-branes) cosmological const
$\L = 4\pi\b^2$. 
%\end{itemize}

%%%%%%%%%%%%%%%%%%%%%%%%%%%%%%%%%%%%%%%%%%%%%%%%%%%%%%%%%%%%%
\subsection{Lightlike Braneworld}
This is a solution with a bulk $D\!=\!5$ space-time consisting of two identical copies 
of the exterior region of $D\!=\!5$ % Anti-de-Sitter-Schwarzschild 
AdS-Schwarzschild  black hole beyond the horizon $r_0$
(``left'' universe for $\eta <0$ and ``right'' universe for  $\eta >0$) 
glued together by a lightlike 3-brane with flat 4-dim world-volume located at
the horizon ($\eta = 0$):
\br
ds^2 = - A(\eta) dv^2 + 2dv\,d\eta + K (r_0 + |\eta|)^2 d\vec{x}^2 \; ,
\lab{AdS5-BW-1}\\
A(\eta) = K (r_0 + |\eta|)^2 - \frac{m}{(r_0 + |\eta|)^2}
\lab{AdS5-BW-2} %\\
% A(0) = 0 \quad ,\quad A(\eta) >0 \; \mathrm{for}\; \eta \neq 0
% \nonu
\er
with $A(0) = 0$ and $A(\eta) >0$ for $\eta \neq 0$, where $\L = - 6 K$ is the bare
$D\!=\!5$ cosmological constant.

The bulk space-time parameters $(K,m)$ are related to the LL-brane parameters
$(T,b_0)$ as: $T^2 = 3K/8\pi^2$ and $b_0 = \frac{2}{3}\sqrt{Km}$.

Because of the shape of the ``effective potential'' $A(\eta)$ \rf{AdS5-BW-2}
a traveling observer along the extra 5-th dimension will ``shuttle'' between the 
two ``universes'' crossing in either direction the $D\!=\!4$ braneworld within 
{\em finite} intervals of his/her proper time.

%%%%%%%%%%%%%%%%%%%%%%%%%%%%%%%%%%%%%%%%%%%%%%%%%%%%%%%%%%%%%%%%%%%%%%%%%%%%%%%%%%%%%%%%%%%%%%%%%%%%%%%%%%%%%%%%%%%%%%%%%%
%%%%%%%%%%%%%%%%%%%%%%%%%%%%%%%%%%%%%%%%%%%%%%%%%%%%%%%%%%%%%
% \section{Discussions and Conclusions}
\section{Conclusions}
%
% \subsection{A Note on Traversability}
To conclude let us recapitulate the crucial properties of the dynamics of 
\textsl{LL-branes} interacting with gravity and bulk space-time gauge fields:

(i) ``Horizon straddling'' -- automatic positioning of \textsl{LL-branes} on (one of) 
the horizon(s) of the bulk space-time geometry. 

(ii) Intrinsic nature of the \textsl{LL-brane} tension as an additional 
{\em degree of freedom} unlike the case of standard Nambu-Goto $p$-branes
(where it is a given \textsl{ad hoc} constant), and which might in particular acquire
zero or negative values.
 
(iii) The stress-energy tensors of the \textsl{LL-branes} are systematically derived
from the underlying \textsl{LL-brane} world-volume Lagrangian actions and provide the 
appropriate source terms on the r.h.s. of Einstein equations to enable the existence of
consistent non-trivial wormhole-like solutions.

(iv) \textsl{LL-branes} naturally couple to Kalb-Ramond bulk space-time gauge fields
which results in {\em dynamical} generation of space-time varying cosmological constant.
In particular, the latter is responsible for creation of a non-singular black
hole with de Sitter interior region below the horizon.

(v) The above properties of \textsl{LL-branes} trigger spontaneous
compactification/decompactification transitions in the bulk space-time manifold.

Further explicit solutions describing multi-``throat'' wormhole-like
space-times of the form ``BR $|$ SdS $|$ SdS $|$ BR'', ``BR $|$ SdS $|$ Schw'', 
``Cyclic'' SdS, as well as ``flat Minkowski $|$ AdS-RN''
will appear in a subsequent paper.

%%%%%%%%%%%%%%%%%%%%%%%%%%%%%%%%%%%%%%%%%%%%%%%%%%%%%%%%%%%%%
%%%%%%%%%%%%%%%%%%%%%%%%%%%%%%%%%%%%%%%%%%%%%%%%%%%%%%%%%%%%%
{\small
\section*{Appendix}

Let us consider for simplicity the \textsl{LL-brane} Polyakov-type action 
\rf{LL-action-dual} for $p=0$, \textsl{i.e.}, the case of {\em lightlike}
(\textsl{LL-}) particle:
\be
S_{\rm LL-particle} = \h \int d\t T b_0^{-\h}
\llb \frac{1}{e} \Bigl(\Xdot^2 - \eps \frac{\udot^2}{T^2}\Bigr) 
- \eps b_0 e\rrb \; ,
\lab{LL-particle}
\ee
where $\Xdot^2 \equiv \Xdot^\m\!\!G_{\m\n}\!\!\Xdot^\n $ and
$e$ is the einbein ($\g_{00} = - e^2\; ,\; \sqrt{-\g} = e$). We will
show that the \text{LL-particle} \rf{LL-particle} is dynamically equivalent to the
standard {\em massless} particle described by the action \rf{test-particle}
with $m_0 = 0$.
% \be
% S_{\rm massless-particle} = \int d\t \frac{1}{2e}\Xdot^2 \; .
% \lab{}
% \ee

Indeed, the action \rf{LL-particle} produces the following equations of motion
w.r.t. $e$, $T$, $u$ and $X^\m$:
\br
\Xdot^2 + \eps \Bigl( b_0 e^2 - \frac{\udot^2}{T^2}\Bigr) = 0 \quad ,\quad
\Xdot^2 - \eps \Bigl( b_0 e^2 - \frac{\udot^2}{T^2}\Bigr) = 0 \; ,
\lab{LL-part-1} \\
\pa_\t \Bigl( \frac{\udot}{eT}\Bigr) = 0 \quad ,\quad
\pa_\t \Bigl(\frac{T}{e}\Xdot^\m\Bigr) + 
\frac{T}{e} \Xdot^\n \Xdot^\l \G^\m_{\n\l} = 0  \; .
\lab{LL-part-2}
\er
Eqs.\rf{LL-part-1} imply ~$\Xdot^2 = 0$ and $e^2 b_0 = \udot^2/T^2$,
% \be 
% \Xdot^2 = 0 \quad ,\quad e^2 b_0 = \frac{T^2}{\udot^2}
% \lab{LL-part-eqs}
% \ee
where the first expression is the standard massless constraint following
from the standard action \rf{test-particle} (with $m_0=0$) upon varying w.r.t.
$e$, whereas the second relation % \rf{LL-part-eqs} 
makes the first Eq.\rf{LL-part-2} an identity. 
The last Eq.\rf{LL-part-2} is obviously equivalent to the standard geodesic 
equation up to a world-line $\t$-reparametrization.

Within the canonical Hamiltonian approach, introducing the canonical momenta
(using the short-hand notation ${\wti e} \equiv e b_0^{1/2}$)
$P_\m = \frac{T}{{\wti e}} G_{\m\n}\!\!\Xdot^\n$ and 
$p_u \equiv - \frac{\eps}{{\wti e}T}\!\udot$ we obtain the canonical Hamiltonian:
\be
H_c = \frac{{\wti e}}{2T} P^2 - \eps \frac{{\wti e}T}{2} \( p_u^2 -1\)
\quad ,\quad P^2 \equiv P_\m G^{\m\n} P_\n \; .
\lab{LL-particle-hamilton}
\ee
Preservation of the primary constraints $p_e = 0$ and $p_T = 0$ (vanishing
canonical momenta of $e$ and $T$) by \rf{LL-particle-hamilton} yields the 
secondary first-class constraints:
\be
P^2 = 0 \quad ,\quad p_u^2 -1 = 0 \; .
\lab{LL-particle-constr}
\ee
Thus, we deduce that $e,T,u$ are non-propagating ``pure-gauge'' degrees of
freedom and we are left with the first relation \rf{LL-particle-constr} which is 
the standard canonical massless constraint resulting from the standard action 
\rf{test-particle} (with $m_0 =0$) within the Hamiltonian formalism. 
}
%%%%%%%%%%%%%%%%%%%%%%%%%%%%%%%%%%%%%%%%%%%%%%%%%%%%%%%%%%%%%
{\small
\section*{Acknowledgments}
E.N. is sincerely grateful to Prof. Branko Dragovich and the organizers
of the Sixth Meeting in Modern Mathematical Physics (Belgrade, Sept 2010) 
for cordial hospitality.
E.N. and S.P. are supported by Bulgarian NSF grant \textsl{DO 02-257}.
% E.G. thanks the astrophysics and cosmology group at PUCV, Chile,  for hospitality.
Also, all of us acknowledge support of our collaboration through the exchange
agreement between the Ben-Gurion University of the Negev and the Bulgarian Academy 
of Sciences.
}
%%%%%%%%%%%%%%%%%%%%%%%%%%%%%%%%%%%%%%%%%%%%%%%%%%%%%%%%%%%%%
%%%%%%%%%%%%%%%%%%%%%%%%%%%%%%%%%%%%%%%%%%%%%%%%%%%%%%%%%%%%%

\end{document}